\documentclass[usenatbib]{mn2e}

\newcommand{\s}{$_{\rm s}$}
\newcommand{\kms}{km~s$^{-1}$}

\newcommand{\etal}{{\it et al.}}
\newcommand{\ie}{{\it i.e.}}

\newcommand{\be}{\begin{equation}}
\newcommand{\ee}{\end{equation}}

\newcommand{\kmsMpc}{km~s$^{-1}$ Mpc$^{-1}$}

\newcommand{\nobs}{1194}

\newcommand{\npoor}{243}
\newcommand{ \ngood}{484}
\newcommand{ \nsnr}{554}
\newcommand{ \nlimit}{465}
\newcommand{ \ndetect}{727}

\newcommand{\mnras}{MNRAS}
\newcommand{\apj}{ApJ}
\newcommand{\apjl}{ApJL}
\newcommand{\aj}{AJ}
\newcommand{\pasp}{PASP}
\newcommand{\aaps}{A\&AS}
\newcommand{\aap}{A\&A}
\newcommand{\apjs}{ApJS}

\usepackage{rotating}
\usepackage{amssymb}

\begin{document}

\title[HI 21cm GBT observations of \nobs ~galaxies]{2MTF III. HI 21cm observations of \nobs ~spiral galaxies with the Green Bank Telescope.}
\author[Karen L. Masters \etal]{Karen L. Masters$^{1,2}$, Aidan Crook$^3$, Tao Hong$^{4,5,6}$, T H. Jarrett$^7$, B\"{a}rbel S. Koribalski$^{8}$, \newauthor Lucas Macri$^9$, Christopher M. Springob$^{5,6,10}$ \& Lister Staveley-Smith $^{5,6}$ \\
$^1$Institute of Cosmology and Gravitation, University of Portsmouth, Dennis Sciama Building, Burnaby Road, Portsmouth, PO1 3FX, UK\\
$^2$South East Physics Network, www.sepnet.ac.uk\\
$^3$Microsoft Corporation, 1 Microsoft Way, Redmond, WA 98052, USA\\
$^4$National Astronomical Observatories, Chinese Academy of Sciences, 20A Datun Road, Chaoyang District, Beijing 100012, China\\
$^5$International Centre for Radio Astronomy Research, M468, University of Western Australia, 35 Stirling Highway, Crawley, WA 6009, \\ Australia\\
$^6$ARC Centre of Excellence for All-sky Astrophysics (CAASTRO)\\
$^7$Astronomy Department, University of Cape Town, Private Bag X3, Rondebosch 7701, South Africa\\
$^8$CSIRO Astronomy \& Space Science, Australia Telescope National Facility, PO Box 76, Epping, NSW 1710, Australia\\
$^9$George P. and Cynthia Woods Mitchell Institute for Fundamental Physics and Astronomy, Department of Physics and Astronomy, \\ Texas A\&M University, 4242 TAMU, College Station, TX 77843, USA\\
$^{10}$Australian Astronomical Observatory, PO Box 915, North Ryde, NSW 1670, Australia\\
\\
{\tt Email: karen.masters@port.ac.uk}
}

\date{MNRAS in press, (accepted 18th June 2014)}
\pagerange{1--13} \pubyear{2014}

\maketitle

\begin{abstract}
We present HI 21cm observations of $\nobs$ ~galaxies out to a redshift of 10,000 km/s selected as inclined spirals ($i\gtrsim60^\circ$) from the 2MASS Redshift Survey. These observations were carried out at the National Radio Astronomy Observatory Robert C. Byrd Green Bank Telescope (GBT). This observing program is part of the 2MASS Tully-Fisher (2MTF) survey. This project will combine HI widths from these GBT observations with those from further dedicated observing at the Parkes Telescope, from the ALFALFA survey at Arecibo, and $S/N>10$ and spectral resolution, $v_{\rm res} < 10$\kms ~published widths from a variety of telescopes. We will use these HI widths along with 2MASS photometry to estimate Tully-Fisher distances to nearby spirals and investigate the peculiar velocity field of the local Universe.  In this paper we report on detections of neutral hydrogen in emission in \ndetect ~galaxies, and measure good signal-to-noise and symmetric HI global profiles suitable for use in the Tully-Fisher relation in \ngood. 
\end{abstract}


\section{Introduction}
The peculiar velocities of galaxies directly trace the amount and distribution of gravitating matter in the Universe --  in the linear regime, peculiar velocities are simply proportional to the underlying gravity field, so in theory a map of peculiar velocities in the local Universe is directly a map of all the matter in the local Universe and environs.  Observationally however, we can only measure the radial component of a galaxy's peculiar velocity, and even this is challenging as it requires a measurement both of its redshift, and of its redshift independent distance (\ie ~in the local Universe via ~$v_{\rm pec} = v_{\rm obs} - H_0d)$. It is always the measurement of $d$ which introduces the dominant error into peculiar velocity measurements, and these errors can be substantial as we discuss below.

For spiral galaxies, one of the most popular techniques to estimate distances has been the use of the luminosity-line width, or Tully-Fisher relation (Tully \& Fisher 1997). This relation can be used to estimate distances to the majority of spiral galaxies typically with an uncertainty of $\sim15-20\%$ (Masters et al. 2006, Masters, Springob \& Huchra 2008; this error is the sum in quadrature of the intrinsic scatter in TF of $\sim 10\%$ and observational errors, dominated by the error introduced into rotation measures by the need to estimate the inclination of the galaxy from photometry - see Masters et al. 2006, 2008 or Springob et al. 2007 for more details). Other techniques (e.g. measuring the brightness of SN1a or the Tip of the Red Giant branch) can measure more accurate distances, however are much more expensive observationally, or have limited range (e.g. for TRGB individual stars must be resolvable). As well as relatively large individual errors on the distances used to derive peculiar velocities, the reconstruction of density fields using this technique has struggled with the sparse sampling and uneven sky coverage introduced by observational realities (e.g. the location and availability of telescopes). In particular most surveys so far have been based on optical selection (e.g.  for all sky TF samples such as Mark III, $N \sim 3000$ -- the third and final version of a concatenation of several early samples published by Willick \etal ~1997; the Spiral Field I-band (SFI) survey of $N \sim 2000$ published by Giovanelli \etal ~1997, and it's follow-on, SFI++, $N\sim 5000$ published in Springob \etal ~2007) and so are strongly biased against low Galactic latitude. This is unfortunate as many of the main features observed in local velocity fields lie close to the Galactic plane (e.g. the Hydra-Centaurus-Shapley Supercluster regions). 

The Two Micron All Sky Survey (2MASS) mapped all of the sky in J, H and K\s-bands and its Extended Source Catalogue (XSC; \citealt{TJ00}) includes roughly half a million galaxies to K\s=13.5 mag. The 2MASS Redshift Survey (2MRS; Huchra \etal ~2012) has measured/collected the redshifts of galaxies uniformly selected from the XSC, and covers 91\% of the sky (avoiding only $|b|>5^\circ$, or $|b|>8^\circ$ towards the Galactic bulge) making it the best survey to use to construct a uniform, all-sky, three dimensional map of the local Universe. The final release of 2MRS published in Huchra et al. (2012) contains $\sim$ 45,000 galaxies and is 97.6\% complete to K\s=11.75 mag. This makes 2MRS the densest sampled {\bf all-sky} redshift survey currently available. 

The 2MASS Tully-Fisher (2MTF) Survey was planned (Masters 2008, Masters et al. 2008) as a matched peculiar velocity survey based on targets in 2MRS with a goal of providing significantly more uniform sky coverage than has previously been available, providing a {\it qualitatively} better sample for velocity--density field reconstructions in the local Universe. A greatly reduced Galactic plane gap, thanks to the NIR selection was designed to aid in studies of the ``Great Attractor'' region which crosses the Galactic Plane at $l\sim300^\circ$. The initial 2MTF selection included all galaxies in the 2MRS catalogue with total $K_s$ magitudes $K_s < 11.25$ mag, redshifts of $cz < 10,000$ km/s, and with axis ratio (from J-band) of $b/a<0.5$ (or $i\gtrsim 60^\circ$). 

 2MTF was designed to make use of 2MASS photometry and existing $S/N>10$ and high resolution ($v_{\rm res}<10$ km/s) rotation widths to construct Tully-Fisher (TF) distances for bright inclined spirals in 2MRS. When the observation plan was made in 2006 about 40\% of the target galaxies had such widths. The ALFALFA (Arecibo Legacy Fast Arecibo L-band Feed Array) survey (Giovanelli et al. 2005, Haynes et al. 2011) was already planned to provide high quality HI measurements for all such galaxies in the high Galactic latitude Arecibo sky, so our new HI observations were planned only to fill in areas of sky not covered by this survey.   A template TF relation in the NIR bands from 2MASS was presented for this sample in Masters et al. (2008), and it has also been used to investigate the MIR 3.4$\mu$ m TF relation in Lagattuta et al. (2014). 
 
 In this paper we present new HI observations for \nobs ~galaxies using the Robert C. Byrd Green Bank Radio Telescope (GBT). These observations result in \ndetect~ HI detections and \ngood~ high quality HI widths suitable for use in the 2MTF survey. Hong et al. (2013) presented the results of a companion Parkes Radio Telescope program to fill in observations in the southern region.  In future work we will combine these data to study the peculiar velocities of galaxies in the local universe (e.g. Hong et al. in prep.). 

\section{Observations and Data Reduction}
This paper presents results from 368 hours of GBT time in the 06A, 06B and 06C semesters (Feb 2006 to Feb 2007) as well as 96 hours in GBT08B (Jun-Aug 2008). 

\begin{figure*}
\includegraphics{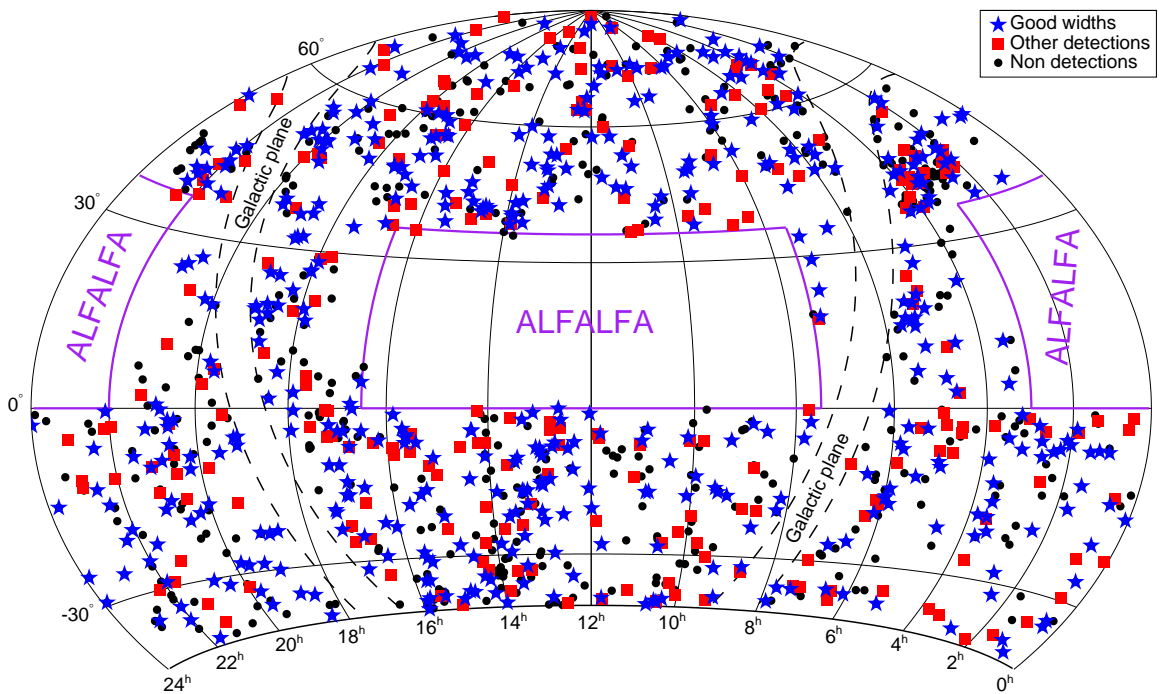}
\caption{Sky distribution of all galaxies observed at the GBT shown on an equatorial projection centred at RA=12h and with RA increasing to the left. Non-detections are shown as black circles, good detections (resulting in HI widths suitable for use in TF; see Section 3.2) as blue stars, and other detections as red squares. Only the part of the sky with $\delta > -40\deg$ (visible from GBT) is shown. The purple lines show the sky areas covered by ALFALFA (Giovanelli et al. 2005), while the dashed lines indicate $|b|=5 \deg$. 
\label{sky}}
\end{figure*}

Our GBT observing strategy was to observe all galaxies making the 2MTF selection (2MRS galaxies with $K_s < 11.25$ mag, $cz < 10,000$ km/s, and $b/a<0.5$, or $i\gtrsim 60^\circ$) which were observable from GB ($\delta > -40^\circ$). We made sure all targets did not already have good ($S/N>10$ and $v_{\rm res}<10$ km/s) HI detections in the literature at the time, and that following visual inspection, they looked like inclined late type spirals. Widths taken from the literature are primarily from the Cornell HI archive, Springob et al. (2005), with some additional widths from Theureau et al. (1998, 2005, 2007); Mathewson et al. (1992) and Paturel et al. (2003). Widths from the HIPASS survey (Barnes et al. 2001) were not used as this survey had $v_{\rm res} = 18$ km/s as well as significantly higher $rms$ (root mean square noise, in signal free parts of the spectum) than we planned for our observations (although HIPASS was invaluable for identifying feasible targets). We also excluded galaxies which were in the sky area of the (then just starting, and now complete) ALFALFA blind HI survey (Giovanelli et al. 2005, Haynes et al. 2011).  Figure \ref{sky} shows the sky distribution of all targeted galaxies in the GBT programme. Hong et al. (2013) describes the results of a similar programme at Parkes to fill in the southern region. 

 Observations at the GBT were done in position switched mode always in pairs of $\sim$5 mins ON/OFF (an initial experiment with frequency switching to save time was not successful as it resulted in baseline structure on similar scale of a typical galaxy rotation width). All galaxies had known redshifts from the 2MRS (Huchra et al. 2012). We used a bandwidth of 12.5 MHz centred on this redshift in the 2006 observations; this was increased to 50 MHz in the 2008 observations. The GBT spectrometer was used with 9 level sampling and 8192 channels. As is standard, the data was saved in 30 second integrations each of which can be separately inspected (we refer to these as ``data segments" below). At 21cm the GBT has a beam size of 9.0\arcmin~ FWHM. 
 
\begin{figure}
%0
\includegraphics[width=9cm]{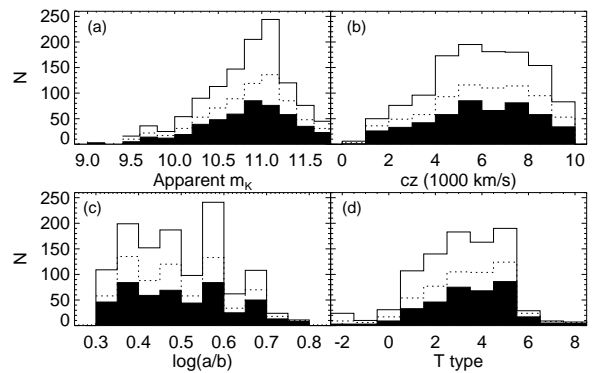}
\caption{The distribution of (a) extinction-corrected apparent $K_s$ magnitude, (b) redshift from 2MRS, (c) J-band axial ratio and (d) T-type as reported by 2MRS for all galaxies observed (upper solid histogram), those detected in HI (dotted line) and those we find to have widths suitable for TF studies (our `` good" detections; filled). \label{sample}}
\end{figure}

 Our observing strategy was designed to obtain high-quality measurements ($S/N \sim10$ at the profile peak) of as many galaxies as possible in the time available. In order to make best use of telescope time, each galaxy was initially observed for a single 5-minute ON/OFF pair. These data were then inspected and if HI was detectable a decision was made by the observer to either continue with repeated scans until reaching $S/N\sim 10$ or to move on to the next nearest galaxy. Our observations are therefore divided into three main categories (1) non-detections in 5-minute pairs ($N= $\nlimit), (2) detections in a single 5-minute pairs (either with $S/N>10$ or such low $S/N$ that it was judged obtaining $S/N\sim10$ was infeasible) and (3) longer observations of multiple 5-minute pairs until $S/N \sim 10$. Histograms showing the extinction-corrected apparent $K_s$ magnitude, redshift, axial ratio and morphological type distribution of all targets, detections, and good detections are shown in Figure \ref{sample}. We only targeted galaxies lacking published high $S/N$ and velocity resolution HI observations, which explains a relative lack of bright, nearby galaxies. 

 All spectra were reduced using the GBTIDL software\footnote{www.gbtidl.nrao.edu}. Radio frequency interference (RFI) was not a major problem except for large signals from the L3 band of the Global Positioning System (GPS) Satellite (at a frequency of 1381 MHz, which corresponds to 21cm emission at about 8000 km/s) which occasionally resulted in a complete loss of several consecutive 30 second data segments. For each galaxy, data segments free of GPS or other major RFI are first combined. Large RFI spikes are then interpolated over before the spectrum is boxcar smoothed (with $n=16$ in the 2006 data or $n=4$ in 2008 data), to give a channel resolution of 5.15 km/s. The spectrum is then Hanning smoothed to an effective resolution of 10.3 km/s. Calibration is performed using the GBT gain curves - good to 2\% in L-band. Baselines are fit to the signal free part of the spectrum. Typically 1st or 2nd order baselines provided an acceptable fit. 
 
 Figures showing profiles of all of our HI detections detections can be found in Appendix A (Figures A.1 and A.2 for ``good" detections and ``other" detections respectively as defined in the next Section), and both raw data and the baseline subtracted spectra are available to download online. \footnote{at the NRAO archive and {\tt icg.port.ac.uk/$\sim$mastersk/2MTF} respectively}

\begin{figure}
%0
\includegraphics[width=8cm]{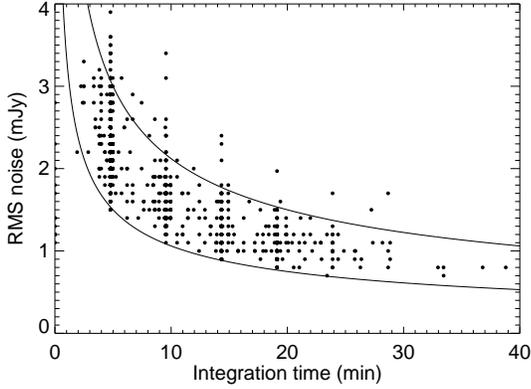}
\caption{$rms$ noise per 5.15 km/s channel (but after Hanning smoothing) vs. integration time for all galaxies observed in this program. The lines show a behaviour of $t_{\rm int}^{-1/2}$ with a factor of two in the zero-point (the upper is for 3mJy $rms$ in 5 mins, the lower for 1.5mJy $rms$ in 5 mins).
\label{tint}}
\end{figure}

Figure \ref{tint} shows the $rms$ noise in our observations as a function of integration time. Our observing strategy based on $\sim$ 5 min ON/OFF pairs is evident in this plot; galaxies with integration times not multiples of $\sim$ 5 mins either had their observations affected by GPS, or other major RFI resulting in a loss of several 30s data segments, or on a small number of cases shorter observations were run to fill the observing schedule. The lines show a behaviour of $t_{\rm int}^{-1/2}$ with a factor of two in the zero-point (the upper is for 3mJy $rms$ in 5 mins, the lower for 1.5mJy $rms$ in 5 mins).

\section{HI Line Fluxes, Velocities and Widths}
 Here we follow Section 3 of \citet{S05} in the corrections which need to be applied to parameters extracted from HI global profiles before physical quantities can be derived from them. This procedure is identical to that used for the 152 HI detections from the Parkes radio telescope presented in Hong et al. (2013). 
 
 \subsection{Integrated Line Flux and Errors}
 Integrated line fluxes are measured from the smoothed and baseline subtracted profiles using an interactive procedure in GBTIDL in which the user marks the part of the spectrum in which HI emission is present. The integrated line flux is then calculated as the total emission within these boundaries.  As described in \citet{S05} this emission must then be corrected for HI self-absorption (pointing offsets and beam attenuation are expected to be negligible at the GBT). An empirically derived correction of $c = (a/b)^{0.12}$ is used (Giovanelli et al. 1994), where $(a/b)$ is the axial ratio of the galaxy, such that $F^c_{\rm HI}= c F_{\rm HI}$. HI masses can then be calculated in the standard way using the formula 
 \be 
 \left(\frac{M_{\rm HI}}{M_\odot}\right) = 2.356 \times 10^5 \left(\frac{D}{\rm Mpc}\right)^2 \left( \frac{F^c_{\rm HI}}{{\rm Jy~km~s}^{-1}} \right)
 \ee (Roberts 1962). 
  
 \subsection{Systemic Velocities and Velocity Widths \label{widths}}
 
Systemic velocities and velocity widths are measured using an adapted version of {\it awv.pro} from the GBTIDL code library. The original code which already allowed for several methods of width measurement, was modified to allow widths to also be measured using a method which fits a polynomial to either side of the profile (between 15-85\% of the peak value (minus $rms$) and finds width values from the fit (see \citet{S05} for a justification of this choice). This modified code has been accepted into future updates of GBTIDL. 

 We report width values for five different methods to allow for easier comparison with other datasets which have in the past used a variety of choices. The methods used are
 \begin{enumerate}
 \item{$W_{\rm F50}$} width measured at 50\% peak-$rms$ on a polynomial fit to both sides of the profile.
 \item{$W_{\rm M50}$} width measured at 50\% of the mean flux value in the profile on the profile itself.
  \item{$W_{\rm P50}$} width measured at 50\% of the peak flux valuer on the profile itself.
  \item{$W_{\rm F20}$} as above but at 20\% peak value.
  \item{$W_{\rm 2P50}$} as above, but at 20\% of the mean of the two peak values.
 \end{enumerate}
 
  We report the systemic velocity of the system as the midpoint of the velocity at 50\% of peak-$rms$. 

\begin{figure}
%0
\centering
\includegraphics[width=8cm]{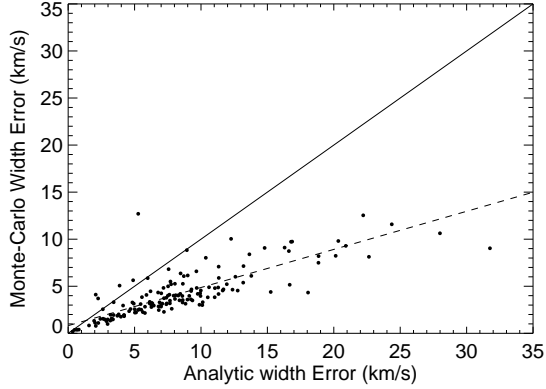}
\caption{A comparison between width errors calculated using a Monte Carlo method (see Hong et al. 2013 for details) and width errors calculated analytically using Equation 1 (in our modified version of {\tt awv.pro}). The solid line shows the one-one relation, while the dashed line is our best fit to these data. \label{eW}}
\end{figure}

 We follow \citet{G97,S05} in selecting $W_{\rm F50}$ as our default width choice since this method reduces the sensitivity of the width measurement on the noise at the edges of the spectrum. and report only this width. The error on this width is calculated as
 \be
 \epsilon_{\rm WF50} =  \sqrt{ (\frac{rms}{a_l})^2 +(\frac{rms}{a_r})^2 },
 \ee
 where $a_l$ and $a_r$ are the slopes fit to the left and right side of the profile respectively, and $rms$ refers to the root mean squared noise per 5.15 km/s channel (after Hanning smoothing) in a signal free part of the spectrum. This differs from the Monte Carlo method presented in Hong et al. (2013); for convenience we present here a conversion formula to convert to Monte Carlo errors: $\epsilon_{\rm MC} = 0.815 + 0.405 \epsilon_ {\rm WF50}$. This conversion is based on 152 galaxies with HI profiles from the Parkes Telescope (Hong et al. 2013) measured in both ways as shown in Figure \ref{eW}.
 
 We also report this width corrected to a final value which accounts for (1) instrumental effects; (2) cosmological broadening; (3) turbulent motions of HI in the disk; and (4) the viewing angle (or inclination, $i$) of the disk. The correction for instrumental effects and cosmology (Eqn 8 in \citet{S05}) is given by
 \be
 W_c = \left [ \frac{W_{\rm F50}-2\Delta v \lambda}{1+z} - \Delta_t \right] \frac{1}{\sin i}, \label{lambda}
 \ee
 where $\Delta v$ is the velocity resolution of the spectrum (5.15 km/s), $z$ is the observed redshift, and $\lambda$ is a factor describing the broadening of the spectrum due to $S/N$ issues. \citet{S05} perform a simulation to discover the impact of instrumental effects on widths with various $\Delta v$ and $S/N$. The values of $\lambda$ we use come from this work and are reported in Table 1. 
   
  We follow \citet{S05} by using a single linear correction for turbulent motions of $\Delta_t = 6.5$ km/s, and finally the inclination, $i$ is estimated using the 2MASS co-added axial ratio $(a/b)$ as
\be
\cos i = \sqrt{\frac{(b/a)^2 - q^2}{1-q^2}},
\ee
where we use $q=0.2$ as a reasonable estimate for the axial ratio of an edge-on spiral (e.g. see Ryden 2004 who report $q=\mu_\gamma\sim0.2$). 

Tables showing these data for our $\ngood$, ``good" detections can be found in Appendix B (Table \ref{tab:data}). By labelling a detection ``good" we indicate that it has $S/N\gtrsim 10$ and in addition has a normal profile shape expected for HI from a single rotating disc galaxy.  We also provide limited data for the low $S/N$ or other ``odd" detections (\ie including those we believe may be contaminated by HI not from the target galaxy), which we do not recommend for use in TF studies (Table \ref{tab:data2}), and $rms$ limits for the non-detections (Table \ref{nondetections}). 

\section{Sample Properties and Notable Detections}

We present in Figure \ref{widths} histograms of the distribution of the observed HI flux and rotation width ($W_{\rm F50}$) of the sample. (Figure 2 showed histograms of the observed K-band magnitude, recessional velocities, axial ratios and morphological types for all targeted and detected galaxies.)

\begin{figure*}
\includegraphics[width=16cm]{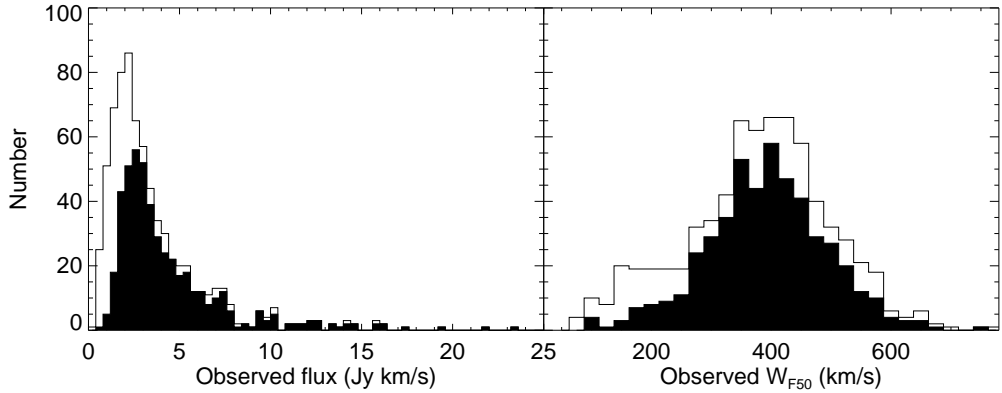}
\caption{Histograms of the observed HI flux, $F_{\rm HI}$ (left) and observed rotation widths, $W_{\rm F50}$ (right) of all HI detected galaxies (unfilled); the $\ngood$ ~well detected are indicated by the filled histogram.
\label{widths}}
\end{figure*}

\begin{figure*}
\includegraphics[width=16cm]{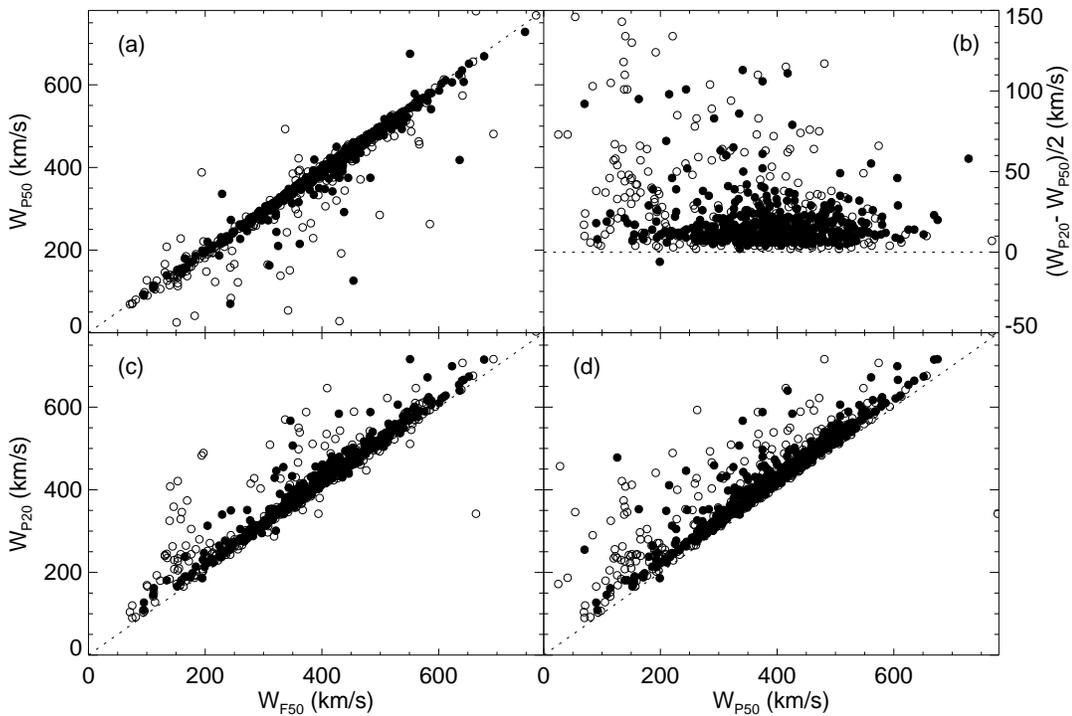}
\caption{Comparison of different width measures for all detections (open circles) and good detections (filled circles). Panel (a) shows $W_{\rm F50}$ against  $W_{\rm P50}$ and (b) shows it against $W_{\rm P20}$. Panel (c) shows $(W_{\rm P20} - W_{\rm P50})/2$ (a measure of the steepness of the sides of the HI profiles), as a function of $W_{\rm P50}$, and (d) shows $W_{\rm P50}$ versus $W_{\rm P20}$. 
\label{widths2}}
\end{figure*}

As described in Section 3.2 we measure widths using five different algorithms. A comparison of the results of three of them is shown in Figure \ref{widths2}. This plots $W_{\rm F50}$ against both (a) $W_{\rm P50}$ and (b) $W_{\rm P20}$, as well as (c) $(W_{\rm P20} - W_{\rm P50})/2$, a measure of the steepness of the sides of the HI profiles, as a function of $W_{\rm P50}$ and finally (d) $W_{\rm P50}$ versus $W_{\rm P20}$. The $\ngood$ well detected galaxies are shown as filled points, while the other detections are shown as open circles. 

\begin{figure}
\centering
\includegraphics[width=8cm,angle=0]{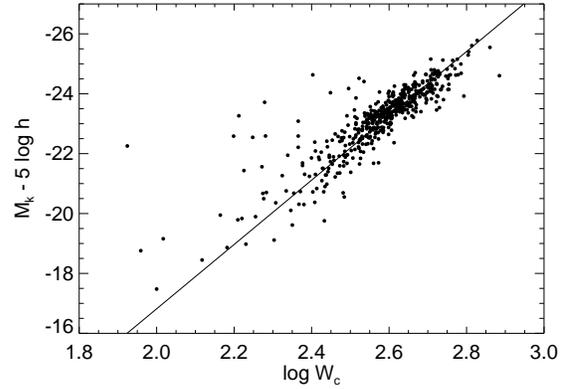}
\caption{The full sample of \ngood ~well detected galaxies shown on the K-band TF relation of Masters et al. (2008).
\label{tf}}
\end{figure}

 As a sample of inclined spirals selected for use with the TF relation, it is reassuring that the $\ngood$ well detected galaxies follow the TF relation well. Figure \ref{tf}  shows these galaxies on the K-band TF relation of Masters et al. (2008; hereafter the 2MTF template relation). The scatter of this sample from the relation is 0.72 mags (28\%). This relatively large scatter is driven by some very large outliers which are all to the upper right of the plot (ie. brighter than expected for their width). Removing 14 galaxies more than 2 magnitudes off the TF relation results in a scatter of 0.48 (19\%). This scatter includes a contribution from peculiar velocities, so should not be interpreted as a measure of the expected distance error for this sample.

\begin{figure}
\includegraphics[width=8cm]{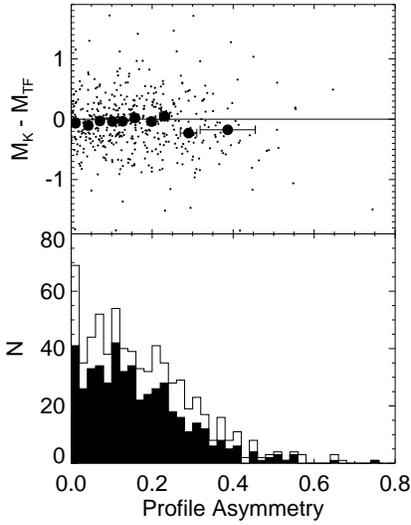}
\caption{Lower: A histogram of the asymmetry (see Eqn 4) of all galaxies (line), with the values for the \ngood ~well detected galaxies shown in the filled histogram. Upper: asymmetry versus TF offset for the \ngood ~well detected galaxies. 
\label{asymmetry}}
\end{figure}

We measure the height of both the right and the left peak for all detections which have two peaks. We define the asymmetry of profiles using these measurements as
\be 
\mathcal{A} = \frac{|P_l - P_r|}{S_p} 
\ee
where $P_l$ is the height of the low velocity peak, and $P_r$ is the height of the high velocity peak, and $S_p$ is the larger of the two.  The lower panel of Figure \ref{asymmetry} shows the distribution of the asymmetry of all detections with two peaks; the high $S/N$ detections are shown in the filled histogram. The median value of this asymmetry is in these ``good" profiles is 14\%, for all profiles with two horns it's 15\%. In the upper panel of Figure \ref{asymmetry} we show the offset from the 2MTF template relation plotted against the profile asymmetry. There is no evidence for any systematic trend, however we do find a suggestion that the galaxies with the most asymmetric profiles have slightly brighter magnitudes than expected for their rotation width (or narrower widths than expected for their magnitude). This may not be a real effect, but could be due to incompleteness bias (e.g. galaxies much dimmer than expected would not appear in the sample, so all large outliers tend to be brighter than the relation). 

\subsection{Fast Rotating Galaxies}

\begin{figure}
\includegraphics[width=9cm]{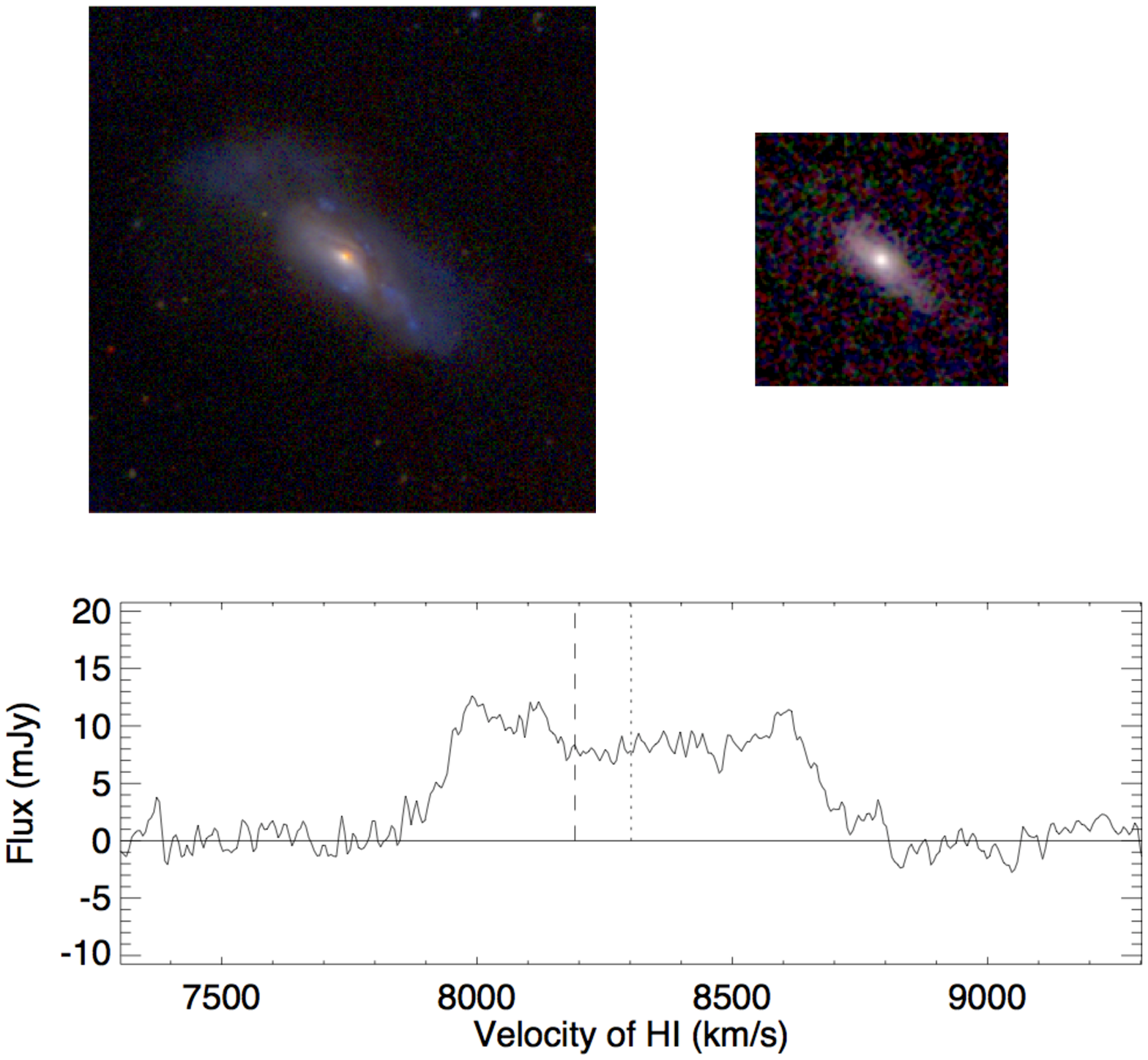}
\caption{The widest observed HI profile in the sample - 2MASXJ 12593965+5320286, or UGC 8107 has a measured width (WF50) of 748km/s. As in Figure \ref{goodwidths}, the catalogue velocity is shown by the dashed line (Huchra et al. 2012), while the dotted line shows the best measured velocity from these data. Above the profile are shown both the SDSS $gri$ image (Lupton et al. 2004), and the 2MASS JHK image (on the same scale; the SDSS image is 4.6\arcmin across, the 2MASS one is 2.3\arcmin across; recall the GBT beam is 9\arcmin ~FWHM). 
\label{widest}}
\end{figure}

There are eleven galaxies in our ``good" sample with widths in excess of 600km/s, implying circular velocities in excess of 300 km/s. The widest profile we observe is that of  2MASXJ 12593965+5320286 (UCG 8107) which is observed with a width of 748km/s (see Figure \ref{widest}). If this value is a measure of the galaxy's rotation, it would make this galaxy one of the fastest rotators known -- spiral galaxies with widths greater than 700 km/s are very rare (Spekkens \& Giovannelli 2006), with the widest known being UGC 12591 with a width of around 1000 km/s (Giovanelli et al. 1986). UGC 8107 has been referred to as a triple system (Berlind et al. 2006) however this was an erroneous identification of a triple due to the SDSS photometric pipeline shredding the galaxy, and the ``phantom galaxies" being assigned the redshift of UGC 8107 in a procedure designed to correct SDSS redshift incompleteness (Berlind priv. com). UGC 8107 is however visibly disturbed in the optical image (Figure \ref{widest}), so while the inferred HI mass (assuming $H_0 = 70$\kmsMpc~ and $D_{\rm Mpc}=v_{\rm HI}/H_0$) of $2.6\times 10^{10} M_\odot$, and its K-band total magnitude of -24.6 are both typical for a galaxy of this width it is possible there are some non-rotational motions being detected. This galaxy is not typical of the 2MTF sample. 

\subsection{Detections offset from Published Redshifts}
\begin{figure*}
\centering
\includegraphics[width=14cm,angle=90]{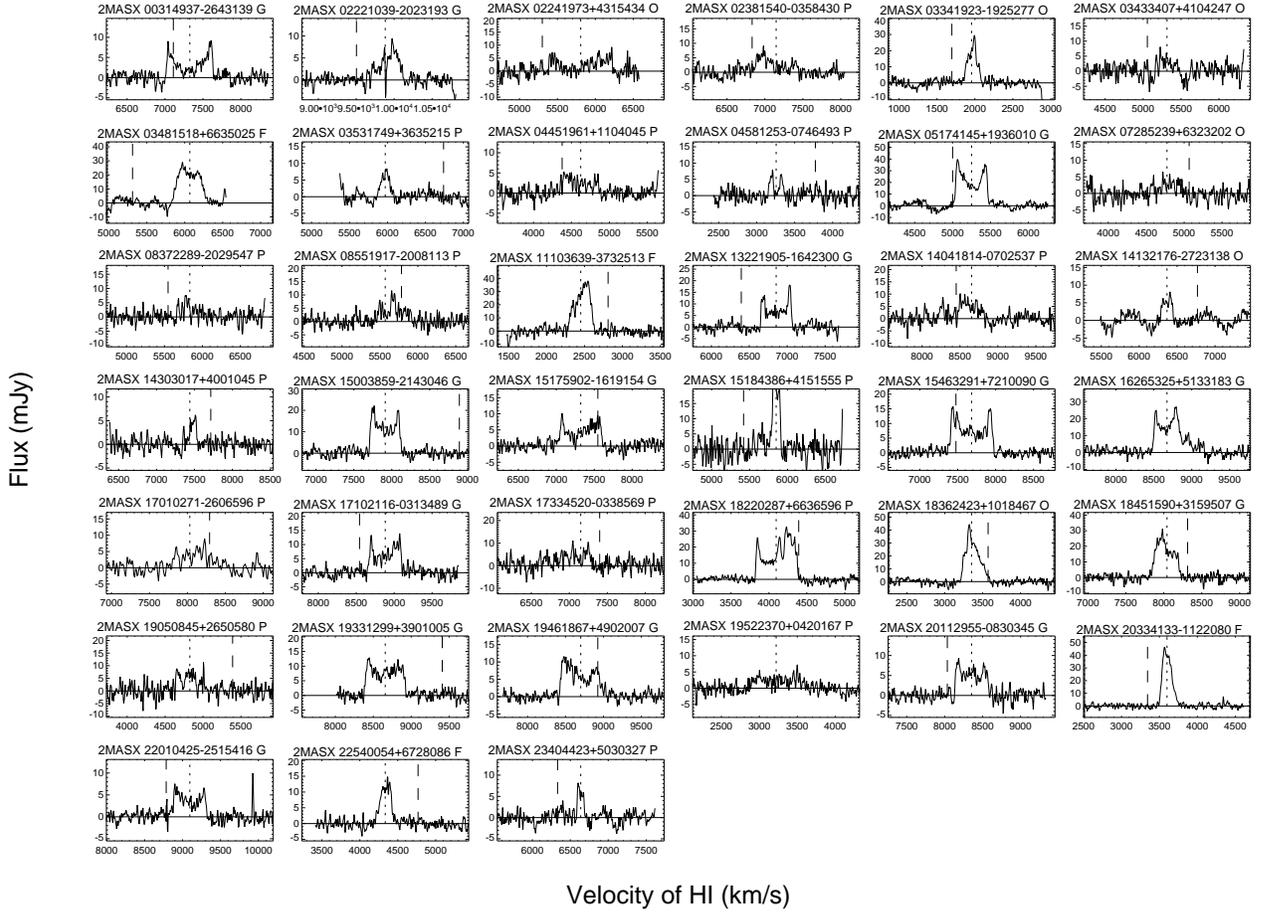}
\caption{Here we present HI profiles for galaxies detected more than 200 km/s from their redshift in 2MRS. Profiles are shown in RA order, with the 2MASXJ ID given. The code following the ID indicates our classification of the detection: 'G' = good, 'F' = fair, 'P' = poor, 'O' = other galaxy (ie. likely HI not from the target galaxy).  The 2MRS velocity (Huchra et al. 2012) is shown by the dashed line where it appears within the plot, while the dotted line shows the best measured velocity from these data. 
\label{offset}}
\end{figure*}

 We list 39 HI detections at recessional velocities more than 200 km/s offset from the redshift value published in 2MRS (Huchra et al. 2012). These galaxies are shown in Figure \ref{offset} and some comments are given below as to possible reasons for the discrepant measurements. The sources coded 'G' (or ``good") are included in the TF sample shown in Figure \ref{tf}.
  
We start with a list of galaxies in which the HI is clearly associated with the target and a previously reported redshift appears to be in error. This is mostly due to reporting low $S/N$ HI detections in which one of the two peaks of the double horned profile was detected, but there is what appears to be an erroneous old optical redshift included in this group also. 

\begin{itemize}
\item 2MASX 00314937-2643139,  $v_{\rm HI} = 7327$ km/s. 2MRS quotes previous HI measurement of 7110 km/s (Thereau et al. 1998) which obviously only picked up low velocity peak.  Because the Nancay beam is so elongated it is quite easy to miss parts of elongated galaxies if they are oriented across the beam. This galaxy also has $v_{\rm opt} = 7247$ km/s from 6dFGS (Jones et al. 2009) 
\item 2MASX 05174145+1936010 ,  $v_{\rm HI} =5250$ km/s. 2MRS reports 5000$\pm$500 km/s from an unpublished Arecibo detection.  There is another measurement of 5255$\pm$5 km/s also from Arecibo (Henning et al. 2010).
\item 2MASX 15175902-1619154,  $v_{\rm HI} = 7322$ km/s. 2MRS reports 7548 km/s from a previous HI measurement (Thereau et al. 1998) which appears to have only detected the high velocity peak. Also 7299 km/s from 6dFGS (Jones et al. 2009).
\item 2MASX 15463291+7210090,  $v_{\rm HI} =7683 $ km/s. 2MRS reports  7477 km/s from a previous HI measurement (Thereau et al. 1998) which appears to have only detected the low velocity peak. 
\item 2MASX 16265325+5133183,  $v_{\rm HI} =8674$ km/s. 2MRS reports 6185 km/s from a re-measurement of optical redshift of 6400km/s cited as data from Arkhipova \& Esipov (1979).
\end{itemize}

The next set of galaxies already have multiple redshifts (either from optical, or low $S/N$ HI) in the literature. Our HI detection provides reasons to prefer one of these over the other. 
\begin{itemize}
\item 2MASX 02221039-2023193,  $v_{\rm HI} =  9883 $ km/s. 2MRS gives 9503 km/s from an redshift from QDOT (Lawrence et al. 1999). There is also a measurement of 9972 km/s (da Costa et al. 1998) which is much closer to the HI detection.
\item 2MASX 02381540-0358430,  $v_{\rm HI} =  7148$ km/s.  Previous redshifts are: 6831 km/s optical Southern Redshift Survey (da Costa et al. 1998), and 6767 km/s from 6dFGS (Jones et al. 2009) .
\item 2MASX 08372289-2029547,  $v_{\rm HI} =5834 $ km/s. 2MRS reports 5547 km/s, also 5612 km/s from 6dFGS (Jones et al. 2009). 
\item 2MASX 08551917-2008113 ,  $v_{\rm HI} =5570$ km/s. 2MRS reports 5784 km/s, also 5633 km/s from 6dFGS (Jones et al. 2009). 
\item 2MASX 15003859-2143046,  $v_{\rm HI} =7914 $ km/s. 2MRS reports 8891 km/s from an optical spectrum (Mathewson \& Ford 1996), there is a measurement of 7876 km/s from 6dFGS (Jones et al. 2009) suggesting this is the better result. 
\item 2MASX 17102116-0313489 ,  $v_{\rm HI} =8883 $ km/s. 2MRS measures 8544 km/s. The measurement of 8828 km/s from 6dFGS (Jones et al. 2009) matched this HI detection better. 
\item 2MASX 22010425-2515416 ,  $v_{\rm HI} =9096 $ km/s. 2MRS reports 8784 km/s from optical (da Costa et al. 1998), however 9083 km/s from 6dFGS (Jones et al. 2009) seems more likely. 
\end{itemize}

We had a few cases in which a profile was detected, but there is a candidate for the source of HI other than the intended target in the 9\arcmin ~FWHM beam of the GBT. In these cases the HI redshift reported here is likely not the correct redshift. These galaxies are not included in the list of good detections even if the $S/N$ is high. 
\begin{itemize}
\item 2MASXJ 02241973+4315434 observed at $v_{\rm 2MRS}=5301$  km/s, detected at $v_{\rm HI}= 5807$ km/s.  Likely UGC 01849 at 6184 km/s in the beam.
\item 2MASXJ 03433407+4104247 observed at $v_{\rm 2MRS}=5056$ km/s,  detected at $v_{\rm HI}= 5315$ km/s.   Likely 2MFGC 03090 at 5333km/s.
\item 2MASXJ 07285239+6323202 observed at $v_{\rm 2MRS}=5066$ km/s, detected at  $v_{\rm HI}=  4774$ km/s. Detection is UGC3850 at 4709km/s, 9\arcmin ~away
\item 2MASXJ 14132176-2723138 (ESO511-G013) observed at $v_{\rm 2MRS}= 6768$ km/s, detected at    $v_{\rm HI}  6367$ km/s. Could be HI from ESO511-G012 at 6341km/s, 6.3\arcmin away.
\item 2MASXJ 18362423+1018467 $v_{\rm 2MRS} = 3571$ km/s. Detection at 3353 km/s  could be part of UGC 11293 (which is 8.9\arcmin from the target). UGC 11293 was detected at 3474km/s by Arecibo (Springob et al. 2005). 
\item 2MASX 18451590+3159507, $V_{\rm HI} =8042 $ km/s. 2MRS measures 8314 km/s. HI might be 2MASXJ J18451354+3157347 (7983 km/s; 2\arcmin), however this is included in the list of good detections. 
\end{itemize}

In the rest of these targets with large offsets between the 2MRS velocity and HI detection, the choice between the published (usually optical) redshift and our HI detection based redshift is not clear.

\subsection{Multiple Detections in Pointing}

\begin{figure*}
%0
\includegraphics[width=18cm]{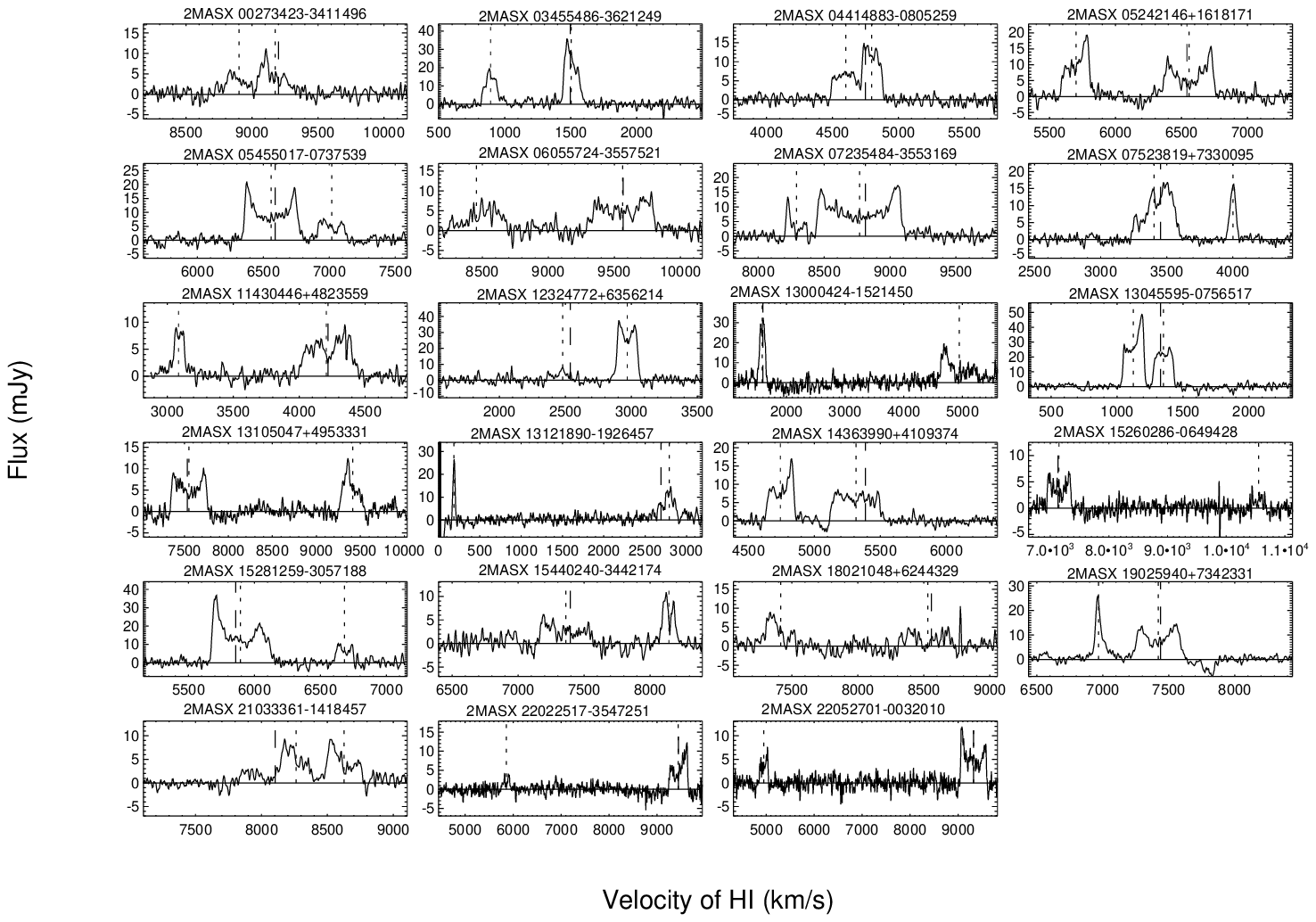}
\caption{Observations in which two or more HI profiles were found. The catalogue velocity is shown by the dashed line (Huchra et al. 2012), while the dotted line shows the best measured velocities of HI profiles in these data. 
\label{2profiles}}
\end{figure*}

We report on serendipitous second profiles detected in the same pointing as the target galaxy. Table \ref{second} lists all observations with multiple HI detections, giving the main velocity, secondary velocity and comments.  Figure \ref{2profiles} shows all the profiles with two detections present. 

\begin{table*}
\caption{Pointings with multiple detections \label{second}}
\begin{tabular}{lcccl} 
2MASXJ ID & $V_{\rm 2MRS}$ & $V_{HI}$  & 2nd $V_{HI}$ & Comments \\ 
 &km/s &km/s &km/s &  \\
\hline\hline
00273423-3411496 & 9200 & 9176 & 8901 & Target ESO 350- G 019 at 9200km/s (2MRS); 9098km/s (6dF). \\ 
& & & & 2MASXJ J00272736-3404353 at cz=9183 km/s, 7\arcmin.\\
03455486-3621249 & 1497  &1503 &  892   \\                                                                         
04414883-0805259 & 4750  & 4797 & 4601 & Confused  \\
05242146+1618171 & 6542 & 6557 & 5701 & 2MASXJ 05244017+1623072, cz=5699km/s, 6.6' from beam\\
05455017-0737539  & 6591 & 6560& 7020\\
06055724-3557521 & 9567  &9563 &   8455 \\                                                                        
07235484-3553169 & 8814  & 8769 &   8292 \\                                                                    
07523819+7330095 & 3454  & 3405 & 4002  \\                                                                                                                                
11430446+4823559 & 4219 & 4206 & 3086 &  \\
12324772+6356214& 2538  &2480 &  2969  & Main galaxy poor detection. Second is likely UGC7700 at 2971km/s.\\  
13000424-1521450 & 1598 & 1588 & 4948 & \\
13045595-0756517 & 1330 & 1352 & 1124 & NGC 4948 at 1330km/s and DDO 163 at 1123km/s \\
& & & & HIPASS detection at 1123km/s probably is a combination of both profiles \\
13105047+4953331 & 7533 & 7552 & 9413 & 2MASXJ J13103350+4953021 at $cz=9417$ km/s in beam.\\ 
13121890-1926457  & 2691 & 2792 & 188 & \\
14363990+4109374& 5387  & 5316& 4742  \\                                                                                          
15260286-0649428 & 7131 & 7149 &  10552 \\
15281259-3057188 &   5858  & 5893 & 6682 \\                                                                                         
15440240-3442174 & 7396 & 7361& 8143 \\
18021048+6244329 & 8558  & 8531 & 7400 \\
19025940+7342331 & 7438  & 7420 &  6968 \\
21033361-1418457 & 8105  & 8263 & 8627 &  2MASXJ J21033437-1425537 (spiral, K=11.5 mag) at $v = 8413$ km/s,  7\arcmin\\
& & & & MCG -02-53-023 also spiral (K=11.3 mag) at $v=8769$ km/s, 8\arcmin.\\
22022517-3547251 & 9443 & 9440 &  5859 & \\
22052701-0032010 & 9321  & 9324 & 4951 & SDSS J220542.16-003340.6 4\arcmin from pointing at 4954 km/s. \\
\hline
\end{tabular}
\end{table*}

The galaxy 2MASX J20331096-0800228 was observed at two different redshifts, because 2MRS reported a redshift of 6273 km/s from 6dF (Jones et al. 2006), while HIPASS detected the galaxy at $3724\pm10$ km/s (Meyer et al. 2004). We detected an oddly shaped HI profile at 3727 km/s which matches the HIPASS detection, and also report a possible marginal detection of HI at 6141km/s.  Both profiles can be seen in Figure \ref{poorwidths}. 

\subsection{Erroneous previous HI detections}

 In three cases we planned to remeasure older HI detections at a higher $S/N$, but did not re-detect the galaxy. This is not unexpected for detections published at low $S/N$. 

\begin{itemize}
\item {\it 2MASXJ 02485298+5302143} has a HI detection published in \citep{RH91} at 4954 km/s which is the only published redshift in NED, however we do not detect this galaxy to $rms=2.1$mJy at this redshift.

\item {\it UGC 4241} (2MASXJ 08092376+5745474) NED lists a redshift of 1235$\pm7$km/s for this galaxy and cites the UZC \citep{F99}. This measurement comes from a published HI observation of this galaxy \citep{RH91} which a published peak flux of $S_p=55\pm5$mJy and measured width of $W_{50}=150$km/s. We observed at this redshift and do not confirm this detection, instead claiming a non-detection to an $rms$ of 2.2mJy. This redshift in NED should therefore be considered incorrect. We re-observed at a suggested redshift of 7633+-23km/s as measured by 2MRS (Huchra et al. 2012) but also claim a non-detection at that redshift (to rms of 3mJy).  

\item 2MASX 13154852-1631080 (NGC 5047) has a HI detection ($S/N=4.9$) at 6330 km/s~ in \citet{RH87}. We report a non-detection to $rms\sim2.3$mJy at this redshift, and also re-observe at the galaxy's optical redshift of 1984 km/s (da Costa etal 1998; 2MRS now reports $v_{\rm opt}1881$ km/s from Mendel et al. 2008 published after these observations occured) for a second non-detection. 

\end{itemize}

\section{Summary}

In this paper we report on HI observations of \nobs~ galaxies selected as highly inclined late-type discs from the 2MASS Redshift Survey (Huchra et al. 2012). These observations, which happened at the Robert C. Byrd Green Bank Telescope in observing seasons 2006A, B, C and 2008B, took a total of 464 hours of GBT time and results in HI detections of \ndetect ~ galaxies with HI widths we believe are useful for Tully-Fisher studies in \ngood. These galaxies are all in parts of the sky inaccessible to the Arecibo radio telescope so contributed to an improved uniformity of HI detections over the sky. We publish the reduced profiles and measured widths with this paper, as well as showing some sample properties and highlighting some of the more interesting detections. 

\paragraph*{ACKNOWLEDGEMENTS.} 

The authors wish particularly to acknowledge John Huchra (1948--2010), without whose vision 2MTF would never have happened. The 2MTF survey was initiated while KLM was a post-doc working with John at Harvard, and its design owes much to his advice and insight. This work was supported by NSF grant AST-0406906 to PI John Huchra. 

The National Radio Astronomy Observatory is a facility of the National Science Foundation operated under cooperative agreement by Associated Universities, Inc. This work is based on observing projects GBT06A-027,  GBT06B-021, GBT06C-049, GBT08B-003: ``Mapping Mass in the Nearby Universe with 2MASS", PI Karen L. Masters. 

 We wish to acknowledge the contributions of Nicholas Reshetnikov, a Harvard undergraduate in 2007 who helped with data reduction of the GBT observations, but who we were unable to locate to invite to be a co-author.

This research has made use of the NASA/IPAC Extragalactic Database (NED) which is operated by the Jet Propulsion Laboratory, California Institute of Technology, under contract with the National Aeronautics and Space Administration. 

The spectra of all HI detections published in this work can be found at http://ict.icrar.org/2MTF/ and http://icg.port.ac.uk/$\sim$mastersk/2MTF/ and through this have been made available to NED for open access download. In addition the raw data is publicly available via the NRAO Data Archive at https://archive.nrao.edu

\appendix
\section{Profile Figures}

\begin{figure*}
\centering
\includegraphics[width=11cm,angle=90]{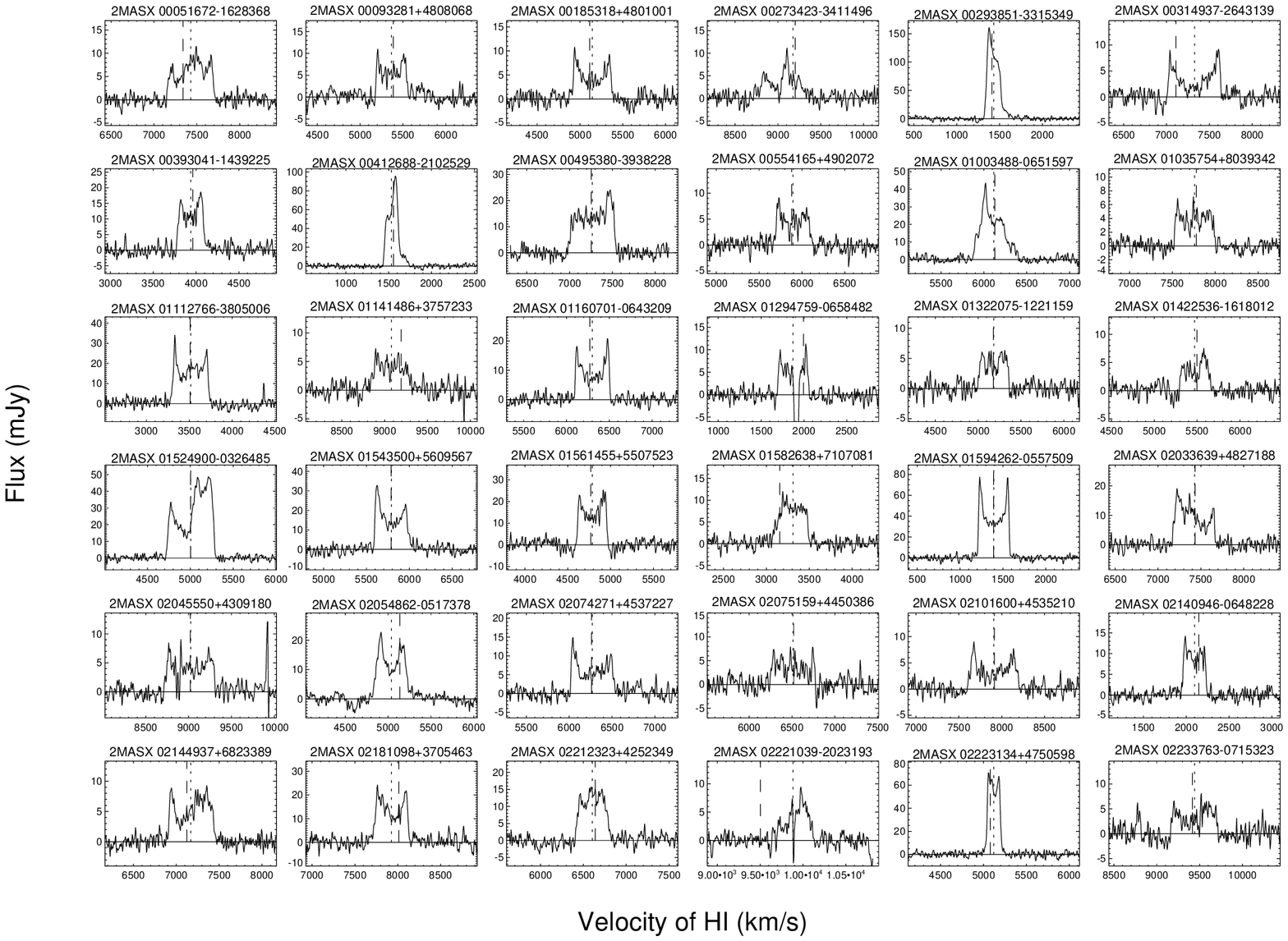}
\caption{Here we present HI profiles for all \ngood~ well detected galaxies (i.e. ~having $S/N \sim 10$ or higher and not too confused or disturbed for use in TF). Profiles are shown in RA order, with the 2MASXJ ID given. The catalogue velocity is shown by the dashed line (Huchra et al. 2012), while the dotted line shows the best measured velocity from these data. This figure is available in its entirety online (7 pages). The first part is shown here for guidance regarding its form and content.
\label{goodwidths}}
\end{figure*}

\begin{figure*}
\centering
\includegraphics[width=11cm,angle=90]{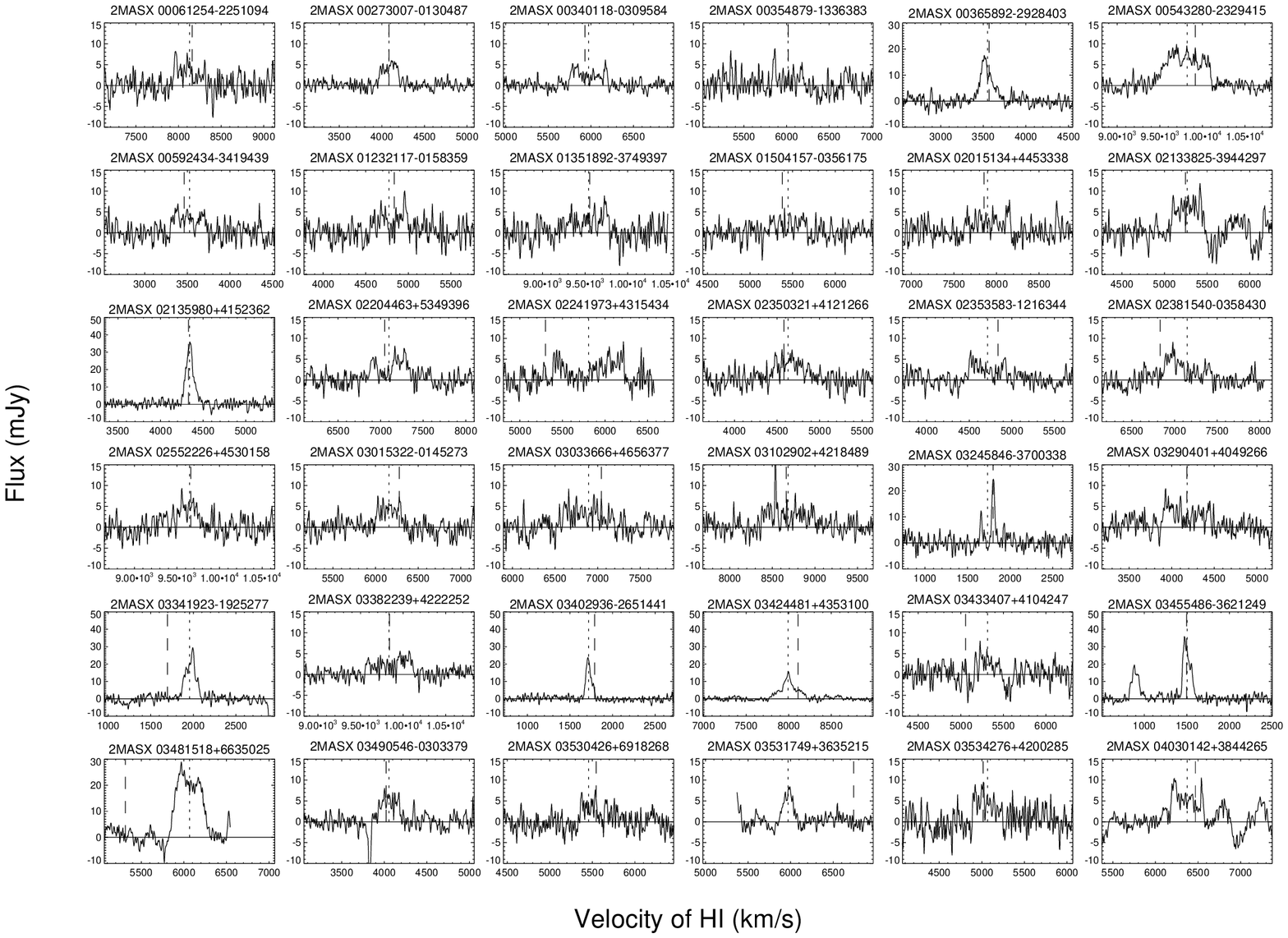}
\caption{Here we present HI profiles for all 243 HI detected galaxies which are not suitable for TF (i.e. having $S/N < 10$, or very confused or disturbed/assymmetric profiles). Profiles are shown in RA order, with the 2MASXJ ID given. The catalogue velocity is shown by the dashed line (Huchra et al. 2012), while the dotted line shows the best measured velocity from these data. This figure is available in its entirety online (4 pages), note that it includes 244 profiles as one galaxy is shown twice. The first page is shown here for guidance regarding its form and content.
\label{poorwidths}}
\end{figure*}

In this appendix we show plots of all HI detections reported here. Figure \ref{goodwidths} shows the profiles for the $\ngood$ good detections (\ie ~$S/N>10$ and not confused or highly asymmetric), while Figure \ref{poorwidths} shows the profiles for all other detections. 

\section{Data Tables}

Here we present raw and corrected parameters for our HI detections as well as limits for the galaxies which were undetected at 21cm. In total we observed \nobs ~galaxies, and detected HI in \ndetect ~(60\%). Non-detections to our $rms$ (typically $\sim 2$ mJy/ km/s for a single observation) are presented for \nlimit~ objects, while in two cases\footnote{2MASX 01254144-3450227 and 2MASX 19595504+4233309} observations were inconclusive due to large baseline errors. Of the galaxies detected in HI, $173$ ~(24\%) were deemed too faint after a single 5 minute observation to be worth following up (\ie it would have taken too long to reach our desired goal of $S/N=10$), the remaining \nsnr ~were followed up after the initial scan if necessary to obtain $S/N\sim10$ to provide reliable widths, and \ngood ~(87\%) of these were found to provide widths suitable for use in TF.
  
 HI profiles for detections useful for TF work (with $S/N > 10$, and also not confused, or hugely disturbed) are shown in Figure \ref{goodwidths}. The parameters are listed in Table \ref{tab:data}\footnote{Only a sample of Table 1 is shown here, the full table is available in ASCII format online} as follows:

\begin{itemize}
\item[(1)] 2MASS XSC ID number
 \item[(2)] Heliocentric redshift from the 2MRS (km/s), see Huchra et al. (2012) for the primary sources.
 \item[(3)] Co-added axial ratio from 2MASS (cite).
 \item[(4)] The morphological type code following the RC3 system. Classification as collated by 2MRS (Huchra et al. 2012). Galaxies with $T=98$ or $T=19$ (meaning unclassified galaxy), were visually inspected and confirmed to be inclined spirals.
 \item[(5-6)] Observed integrated 21cm line flux $F_{\rm HI} = \int S dV$ in Jy km/s, and associated error. 
  \item[(7)]  Heliocentric redshift of the detected 21cm line taken as the midpoint of the 50th\% level of the profile using the $W_{P50}$ method (km/s).
  \item[(8-13)] The width of the 21cm line profile measured five different ways (as discussed in Section \ref{widths}). Widths are $W_{F50}$, $W_{M50}$, $W_{P50}$, $W_{P20}$, $W_{2P50}$. Column (13) is the error on $W_{F50}$. 
 \item[(14-15)] HI line width and its error after making all corrections discussed in Section \ref{widths} (km/s). Note that it is $W_{F50}$ which is corrected.
  \item[(16)] The peak signal-to-noise ratio of the 21cm line, $S/N$.
 \item[(17)] The value of the velocity width instrumental parameter, $\lambda$ (see Eqn. \ref{lambda}). 
 \end{itemize}

\begin{table*}
\tiny
\caption{ ~H\,{\sc i} Parameters of good detections (high $S/N$ and not confused/disturbed)\label{tab:data}}
\begin{tabular}{lrrcccrrrrrrrrccccc}
\hline\hline
2MASXJ  &$V_{\rm 2mrs}$  &$T$ &$b/a$ &$F_{\rm HI}$ &$\epsilon_F$ &$V_{HI}$& $W_{\tiny F50}$ & $W_{\tiny M50}$ & $W_{\tiny 2P50}$ & $W_{\tiny P50}$ & $W_{\tiny P20}$ & $\epsilon_{\tiny WF50}$ & $W_{c}$ & $\epsilon_{Wc}$ & S/N& $\lambda$\\
-~-&\tiny km/s&-~- &-~-&\multicolumn{2}{c}{\tiny Jy~km/s} &\multicolumn{9}{c}{km/s} & -~- & -~-  \\
(1) & (2) & (3) & (4) & (5) & (6) & (7) & (8) & (9) & (10) & (11) & (12) & (13) & (14) & (15) & (16) & (17)  \\
\hline
00051672-1628368	 &	7412	 &6&	0.36	  &3.57	&0.14    &7429	&525	 & 535 & 531 &	502 & 548	 &  10  &	528	&12.1 & 10.5	& 0.337\\	
00093281+4808068	 &	5388	 &2&	0.48	 & 2.27	&0.14   & 5367	&359	 & 373 & 354 &353 & 377	  &  6	  & 384	&11.2&	  8.5	 &0.265	\\
00185318+4801001	  &  	5113	 &4&	0.44	 &2.35	&0.13  & 5144	&445 & 454 & 442 &440 & 457	 & 7	  & 467	&11.6&	  9.8	 &0.314	\\
00273423-3411496	   & 	9200	& 2&	0.30	& 1.35     &0.10   & 9176	&243	 & 255 & 233 &70  & 255	 & 49  &	232	&48.9&	  9.6	& 0.307\\	
00293851-3315349	    &	1455& 3&	0.38	 &21.75	&0.18    &1432	&168 &193 & 166 &156   & 190 &    2  & 166	 &3.6&	 67.1	 &0.398\\
....... \\
\hline\hline
\end{tabular}
\medskip 
Table~\ref{tab:data} is available in its entirety online. A portion is shown here for guidance regarding its form and content.
\end{table*}

Data for all other detections ($N=$\npoor) are presented in Table \ref{tab:data2}. In this case only an estimate of the width is given which has large uncertainty (this is WF50 where that method worked, otherwise it is WP50)\footnote{this table contains 244 lines as one galaxy was observed at two separate central velocities with a weak detection in both bands}. We do not recommend using these widths for science, as they are either based on low $S/N$ data or confused or disturbed profiles. The HI profiles for these galaxies are shown in Figure \ref{poorwidths}

\begin{table*}
\tiny
\caption{ ~H\,{\sc i} Parameters of all other detected galaxies \label{tab:data2}}
\begin{tabular}{lccccccc}
\hline\hline
2MASXJ ID &$V_{\rm 2MRS}$  &$F_{\rm HI}$ &$\epsilon_F$ &$V_{HI}$& $W_{HI}$ & S/N & Comments\\
-~-  &[km/s]& \multicolumn{2}{c}{[Jy~km/s]} &\multicolumn{2}{c}{[km/s]} & -~- \\
(1) & (2) & (3) & (4) & (5) & (6) & (7)  \\
\hline
00061254-2251094&    8160&    1.27&    0.23&        8129&         376&    3.9 & -\\
00273007-0130487& 4084  &1.03 &0.08 &4085  &230  &6.1   & -\\
00340118-0309584&        5929&    1.22&    0.13&        5973&         438&    5.6& -\\
00354879-1336383&        6021&    0.72&    0.25&        6019&         366&    3.9& -\\
00365892-2928403 &3569  &2.49 &0.16 &3551 & 263 & 9.8  &  s\\
00543280-2329415 &9917 & 3.81 &0.15 &9821  &561 & 9.0 &   r\\
00592434-3419439&        3463&    1.40&    0.21&        3526&         404&    3.7& -\\
01232117-0158359&        4835&    1.39&    0.24&        4774&         425&    4.8& -\\
......\\
\hline\hline \\
\end{tabular}

\medskip Table~\ref{tab:data2} is available in full in electronic form. A portion is shown here for guidance regarding content. The comment codes are s=single peaked; c=confused; d=disturbed; b=not all in beam; e=near edge of bandpass; r=significant rfi removed from within profile; nt=not intended target 
\end{table*}

Non detections for 465 galaxies are reported in Table \ref{nondetections}. In this case only the $rms$ noise is reported from the 21cm observations, along with the central velocity of the observation.
Two galaxies are  reported with non-detections at two different redshifts (2MASXJ 08092376+574547 and 2MASXJ 13154852-1631080), so this table contains 467 rows.

\begin{table}
\tiny
\centering
\caption{List of non-detections \label{nondetections}}
\begin{tabular}{lcccc}
\hline\hline
2MASXJ ID  &$V_{\rm 2MRS}$  &$rms$ \\
-~- &km/s & mJy \\
(1) & (2) & (3)  \\
\hline
00110081-1249206  & 6775 &  2.2 \\
00180298-3255454  & 7788 &  2.8 \\
00221510-1435348  & 6715 &  2.2 \\
00475430+6807433  & 1273 &  2.4 \\
00521377+4419514  & 5328 &  2.3\\
.....\\
\hline\hline
\end{tabular}

\medskip Table~\ref{nondetections} is available in electronic form. A portion is shown here for guidance regarding content.
\end{table}

\end{document}